\documentclass{qjmam}
\usepackage{amssymb}
\usepackage{amsmath}
\graphicspath{{./figures/}}
\usepackage{cleveref}
\usepackage{color}
\usepackage{calrsfs}
\usepackage[symbol]{footmisc}

\startpage{1}
\yr{2025}
\vol{78}
\issue{2}

\newcommand{\pfr}[2]{\ensuremath{\frac{\partial #1}{\partial #2}}}
\newcommand{\pfi}[2]{\ensuremath{{\partial #1}/{\partial #2}}}

\newcommand{\Sch}{Sc}
\newcommand{\Ray}{Ra}
\newcommand{\Pra}{Pr}
\newcommand{\Lew}{Le}
\newcommand{\eff}{\mathrm{eff}}
\newcommand{\vect}[1]{\mathbf{#1}}
\newcommand{\ep}{\varepsilon}

\newcommand{\beq}{\begin{equation}}
\newcommand{\eeq}{\end{equation}}

\DeclareMathAlphabet\mathbfcal{OMS}{cmsy}{b}{n}

\DeclareMathAlphabet\mathbit
    \encodingdefault\rmdefault\bfdefault\itdefault
\DeclareOldFontCommand{\bi}{\normalfont\bfseries\itshape}{\mathbit}

\newcommand{\be}{\begin{equation}}
\newcommand{\ee}{\end{equation}}

\def\fakebold#1{\relax\ifvmode\leavevmode\fi%
\ifmmode%
\setbox0=\hbox{$#1$}%
\else%
\setbox0=\hbox{#1}%
\fi%
\kern-.02em\copy0 \kern-\wd0%
\kern .04em\copy0 \kern-\wd0%
\kern-.0125em\raise.02em\box0%
}%

\begin{document}

\title[{Shear-induced force and dispersion in a Hele-Shaw cell}] {Shear-induced force and dispersion due to buoyancy in a horizontal Hele-Shaw cell}

\author[P.~Rajamanickam] {Prabakaran Rajamanickam}

\address{Department of Mathematics and Statistics, University of Strathclyde, Glasgow G1 1XQ, UK}

\received{\recd 29 September 2024 \revd 22 April 2025}

\maketitle

\eqnobysec

\begin{abstract} 
This paper investigates shear flow in a Hele-Shaw cell, driven by varying horizontal buoyancy forces resulting from a horizontal density gradient induced by a scalar field. By employing asymptotic methods and taking the dependence of density and transport coefficients on the scalar field into account, effective two-dimensional hydrodynamic equations coupled with the scalar conservation equation are derived. These equations determine an effective diffusion coefficient for the scalar field accounting  for shear-induced diffusion, and an effective shear-induced  buoyancy force that modifies the classical Darcy's law. The derived equations provide a foundation for future research into various problems involving scalar transport in horizontal Hele-Shaw cells.
\end{abstract}

\section{Introduction}
\label{sec:intro}

When a horizontal density gradient exists within a fluid, mechanical equilibrium is impossible due to varying buoyancy forces acting on fluid elements at different locations but the same altitude~\cite{landau1987fluid}. Such a gradient may develop from a spatial variation of any scalar field, such as the fluid temperature or the concentration of a dispersing solute, upon which the density depends. For instance, in natural-convection (or, indirect-convection) boundary layers, a horizontal pressure gradient develops when there is a horizontal temperature gradient~\cite{stewartson1958free}. An interesting phenomenon  occurs when fluid motion due to buoyancy force occurs in narrow horizontal geometries: Taylor or shear-induced dispersion. This arises due to a local-scale interaction between a strong shear flow and a slowly-varying scalar field, as first elucidated by Taylor in 1953~\cite{taylor1953dispersion}. The local-scale interaction effectively appears  as a diffusion process for the scalar on the large scale. Taylor calculated the effective diffusion coefficient, which was found to be proportional to the square of the flow Peclet number.

In the original Taylor's analysis, the shear flow was driven by an externally imposed pressure gradient. However, it should be noted that the dispersing scalar field itself can induce shear flow  due to the indirect buoyancy forces mentioned above. This fact was first appreciated by Erdogan and Chatwin~\cite{erdogan1967effects,chatwin1976some}, who derived an effective diffusion coefficient for a buoyancy-induced flow in a horizontal pipe. In a two-dimensional Hele-Shaw cell configuration, the effective diffusion coefficient $D_{\eff,x^*}$, say for the $x^*$-direction, for the (non-dimensional) scalar field $\theta$, is given by
\begin{equation}
    D_{\mathrm{eff},x^*}^* = D^* \left[1+\frac{\gamma g^2h^8\alpha^2}{\nu^{*2}D^{*2}}\left(\pfr{\theta}{x^*}\right)^2\right], \label{dimDeff}
\end{equation}
where $D^*$ is the molecular diffusivity of the scalar, $\gamma=2/2835$ is a numerical factor, $g$ is the gravitational acceleration, $\nu^*$ is the kinematic viscosity of the fluid, $h$ is the channel half-width and $\alpha=-\rho^{*-1}\pfi{\rho^*}{\theta}$ is an volumetric expansion coefficient ($\rho^*$ here is the fluid density). This formula, derived under the Boussinesq approximation, leads to a nonlinear diffusion equation for $\theta$, known as the Erdogan--Chatwin equation~\cite{smith1978asymptotic,crowe2018evolution,salmon2020buoyancy}. A simple derivation of the above formula can be found in~\cite{young1991shear}. 

 \begin{figure}
\centering
\includegraphics[scale=0.7]{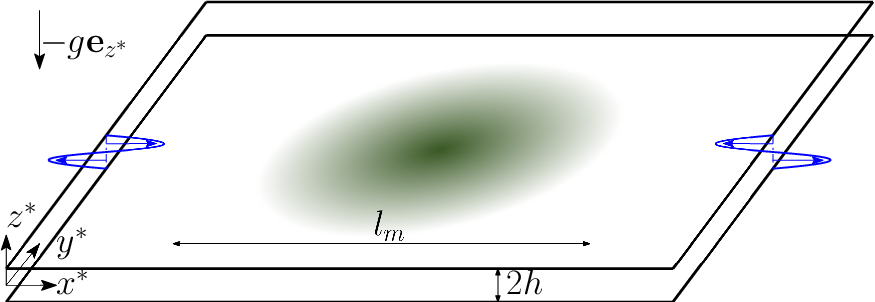}
\caption{Schematic illustration of a scalar field dispersing in a horizontal Hele-Shaw cell.} 
\label{fig:sch}
\end{figure}

This paper aims to generalise the results of Erdogan and Chatwin beyond the Boussinesq approximation, allowing density and all transport coefficients, to depend on the scalar field. By relaxing the Boussinesq approximation, we achieve a more comprehensive coupling between hydrodynamics and the scalar field. The plan of the paper is to obtain the effective two-dimensional depth-averaged governing equations in a Hele-Shaw cell, starting from the full three-dimensional Navier--Stokes equations, by employing the well-known long-time (or 
large-scale) asymptotic method~\cite{chatwin1970approach,linan2020taylor}. The study will uncover an effective buoyancy force appearing in Darcy's law  and an effective diffusion coefficient appearing in the scalar-field equation.

\section{Problem formulation}
\label{sec:formulation}

As mentioned in the introduction, we consider a narrow fluid layer parallel to the $x^*y^*$-plane within a Hele-Shaw cell with a half-width $h$. The gravity vector points along the negative $z^*$-axis, as depicted in Fig.~\ref{fig:sch}. Let the characteristic length, time and velocity scales associated with the scalar field $\theta(x^*,y^*,z^*,t^*)$ be
\begin{equation}
    l_m, \qquad t_m = \frac{l_m^2}{D_c^*}, \qquad U_m = \frac{D_c^*}{l_m} \label{mixing}
\end{equation}
where $D_c^*$ is a reference value for the  diffusion coefficient of the scalar. Generally speaking, shear-induced dispersion arises under two basic limiting conditions. First of all, the mixing length scale $l_m$ must be significantly greater than the channel half-width $h$. Equivalently, the mixing time scale $t_m$ should be greater than the diffusion time $t_h=h^2/D_c^*$ across the channel. This limit is represented by a small parameter $\ep$, which is defined by
\begin{equation}
    \ep= \frac{h}{l_m}=\sqrt{\frac{t_h}{t_m}} \ll 1. \label{epsil}
\end{equation}
Secondly, the shearing motion induced by the buoyancy force must be stronger than the mixing speed $U_m$. The characteristic velocity scale, $U_b$ for a representative relative density difference $\Delta \rho^*/\rho_c^*$, may be defined as
\begin{equation}
    U_b=\frac{\Delta\rho^*}{\rho_c^*}\frac{gh^3}{\nu_c^*l_m}, \label{buoyscale}
\end{equation}
which is obtained by balancing the characteristic horizontal gradient of the hydrostatic pressure gradient, i.e., $(\Delta\rho^* g z)/l_m$, with the viscous force $\rho_c^*\nu_c^* U_b/h^2$. The second limiting condition is then expressed as 
\begin{equation}
    \frac{U_b}{U_m} \sim \frac{1}{\ep} \gg 1. \label{strong}
\end{equation}

The two limits~\eqref{epsil} and~\eqref{strong} simply indicate a strong buoyancy-induced shear flow within a narrow, horizontal fluid layer. To represent the second limit, we introduce a non-dimensional number, say a Rayleigh number $\Ray$, but otherwise also a Prandtl number $\Pra$ (or equivalently a Schmidt number $\Sch$). These numbers are defined by
\begin{equation}
   \Ray = \frac{h}{l_m}\frac{gh^3}{\nu_c^* D_c^*},\qquad \Pra = \frac{\nu_c^*}{D_c^*}.  \label{nondimnumber}
\end{equation}
In defining the Rayleigh number, the ratio $\Delta \rho^*/\rho_c^*$ is excluded as it can be of order one in gaseous flows. However, for liquids, this ratio is typically small and an alternative Rayleigh number $\Ray_l=\ep U_b/U_m= \Ray \Delta\rho^*/\rho_c^*$ can be introduced at a later stage if necessary.

To facilitate the analysis, we introduce the following non-dimensionalization: 
\begin{equation}
\begin{aligned}
    &t = \frac{t^*}{t_m}, \quad (x,y) = \frac{1}{l_m}(x^*,y^*), \quad z=\frac{z^*}{h},  \quad \widehat{\vect v} = \frac{\widehat{\vect v}^*}{U_m}, \quad p = \frac{p^*h^2}{\rho_c^*\nu_c^*D_c^*},  \\
    &\rho = \frac{\rho^*}{\rho_c^*}, \quad 
    \mu = \frac{\rho^* \nu^*}{\rho_c^*\nu_c^*}, \quad \lambda=\frac{\rho^* D^*}{\rho_c^* D_c^*},\label{nondim}
\end{aligned}
\end{equation}
in terms of which the low Mach-number Navier--Stokes equations can be written as
\begin{align}
    \pfr{\rho}{t} +  \widehat\nabla\cdot(\rho \widehat{\vect v}) &=0, \label{3dcont}\\
    \frac{\rho}{\Pra}\left(\pfr{\widehat{\vect v}}{t}+ \widehat{\vect v}\cdot \widehat\nabla\widehat{\vect v}\right) &= -\frac{\widehat\nabla p}{\ep^2} + \widehat\nabla\cdot\boldsymbol\tau - \frac{\Ray}{\ep^4}\rho\, \vect e_z, \label{3dmom}\\
   \rho\left(\pfr{\theta}{t}+ \widehat{\vect v}\cdot \widehat\nabla \theta\right) &= \widehat\nabla\cdot(\lambda \widehat\nabla \theta) + Q(\vect x,t,\theta), \label{thetadim}\\   
    \rho =\rho(\theta), \quad \mu &= \mu(\theta) \quad \lambda=\lambda(\theta), \label{eqnst}
\end{align}
where the viscous stress tensor is defined by
\begin{equation}
    \boldsymbol\tau = \mu\left[\widehat\nabla \widehat{\vect v} + \widehat\nabla \widehat{\vect v}^{T}-\tfrac{2}{3}(\widehat\nabla\cdot\widehat{\vect v})\mathbf{I}\right].
\end{equation}
In these equations, $\widehat\nabla=(\nabla,\partial_z/\ep)$ denotes the three-dimensional gradient operator in which $\nabla=(\partial_x,\partial_y)$ denotes the two-dimensional gradient operator and  $\widehat{\vect v}=(\vect v,w)$ denotes the three-dimensional velocity field in which $\vect v=(u,v)$ denotes the two-dimensional velocity field. Similarly $\widehat{\vect x}=(\vect x,  z)$ is the three-dimensional position vector and $\vect x=(x,y)$ is the two-dimensional position vector. Furthermore, a term $Q(\vect x,t,\theta)$ is included in the scalar equation, to represent any volumetric source or sinks such as chemical reaction, that may be present. Equation~\eqref{eqnst} represents the equation of state and the constitutive relations for the transport coefficients.

Since the fluid is bound between two parallel rigid walls, we have
\begin{align}
    \vect v=\vect 0, \quad w =0\quad \text{at} \quad z=\pm 1 \label{Abc}.
\end{align}
If we permit for small leakages of the scalar field, then we have
\begin{equation}
    -\frac{\lambda}{\ep}\pfr{\theta}{z} =   \pm \ep S^{\pm}(\vect x,t,\theta) \quad \text{at} \quad z=\pm 1 \label{flux}
\end{equation}
where $S^+(\vect x,t,\theta)$ and $S^-(\vect x,t,\theta)$ are prescribed functions; $-(S^++S^-)/2$ measures the average flux of scalar leaking through the walls.

\section{Asymptotic solution in the limit $\ep\to 0$}\label{sec:asym}
The solution to the non-dimensional problem written down in the previous section is developed here in the asymptotic limit $\ep\to 0$, while regarding the non-dimensional numbers~\eqref{nondimnumber} as free order-unity parameters in the problem. The solution is sought in the form of a regular perturbation series
\begin{equation}
    \begin{aligned}
  \vect v = \ep^{-1}\vect v_0(\vect x,z,t) +   \vect v_1(\vect x,z,t)  + \cdots, &\quad  w = 0 +   w_1(\vect x,z,t)  + \cdots, \\
   p = \ep^{-1} p_0(\vect x,z,t)+ p_1(\vect x,z,t)+ \cdots , &\quad
   \theta = \theta_0(\vect x,z,t) + \ep\theta_1(\vect x,z,t)+\cdots,  \label{RPS}
\end{aligned}
\end{equation}
in which the leading-order scaling for $\vect v$ and $p$ indicates the presence of strong buoyancy-induced shear flow. The expansion for density is given by $\rho=\rho_0+\ep\rho_1+\ep^2\rho_2+\cdots$ where 
\begin{equation}
    \rho_0=\rho(\theta_0), \quad \rho_1 = \theta_1\pfr{\rho_0}{\theta_0}, \quad \rho_2=\theta_2 \pfr{\rho_0}{\theta_0} + \frac{\theta_1^2}{2} \pfr{^2\rho_0}{\theta_0^2}, \quad \text{etc.} 
\end{equation}
Similar expansions can be written down for the functions $\mu(\theta)$, $\lambda(\theta)$, $Q(\vect x,t,\theta)$ and $S^\pm(\vect x,t,\theta)$. 

Substituting the expansions~\eqref{RPS} into equations~\eqref{3dcont}-\eqref{flux} and collecting terms of different orders of $\ep$, we obtain a set of equations at different orders, which are solved below. To facilitate further calculations, it is convenient to define the depth-averaged value of a physical variable, say $\varphi$, across the fluid layer, by
\begin{equation}
    \langle\varphi\rangle = \frac{1}{2}\int_{-1}^{+1}\varphi\,dz. \label{avg}
\end{equation}

\subsection{Leading-order problem: Ostroumov flow}\label{sec:ostroumov}
At leading order, we obtain
\begin{equation}
    \nabla\cdot(\rho_0\vect v_0) + \pfr{}{z}(\rho_0 w_1) = 0, \quad  \nabla p_0=\pfr{}{z}\left(\mu_0\pfr{\vect v_0}{z}\right), \quad \pfr{p_0}{z}=-\Ray \rho_0, \quad \pfr{}{z}\left(\lambda_0\pfr{\theta_0}{z}\right)=0  \label{cont0}
\end{equation}
subject to the boundary conditions
\begin{equation}
    \lambda_0\pfr{\theta_0}{z} = 0, \quad \vect v_0 = 0, \quad w_1=0, \quad \text{at} \quad z=\pm 1.
\end{equation}
On the lateral sides, far away from the mixing region, we shall assume that there is no fluid motion. In other words, the fluid motion emerge entirely due to the mixing processes and not by any external means.

Integration of the last equation subject to the condition shows that $\theta_0=\theta_0(\vect x,t)$, $\rho_0=\rho_0(\vect x,t)$, $\mu_0=\mu_0(\vect x,t)$ and $\lambda_0=\lambda_0(\vect x,t)$. Integration of the continuity and the momentum equations leads to a type of solution that was originally studied in~\cite{ostroumov1952svobodnaya,birikh1966thermocapillary,hansen1966gravitational}, in which $\vect v_0$ is a cubic polynomial in $z$. This is due to the horizontal pressure gradient $\nabla p_0$ being proportional to $\nabla \rho_0 z$ which arises from differential hydrostatic balance. Specifically, the solution for our problem is given by\footnote{The reader may also find the dimensional form of the solution useful. It is given by $\pfr{p_*}{x_*}=-g\pfr{\rho_*}{x_*}z_*$, $\pfr{p_*}{y_*}=-g\pfr{\rho_*}{y_*}z_*$, $u_* = \frac{g}{\mu_*}\pfr{\rho_*}{x_*}z_*(h^2-z_*^2)$, $v_* = \frac{g}{\mu_*}\pfr{\rho_*}{y_*}z_*(h^2-z_*^2)$ and $w_*=\frac{g}{24\rho_*}\left[\pfr{}{x_*}\left(\frac{\rho_*}{\mu_*}\pfr{\rho_*}{x_*}\right)+\pfr{}{y_*}\left(\frac{\rho_*}{\mu_*}\pfr{\rho_*}{y_*}\right)\right](h^2-z_*^2)^2$.} 
\begin{equation}
    \nabla p_0 = -\Ray \nabla\rho_0 z, \quad \mu_0\vect v_0 =\frac{\Ray}{6}\nabla\rho_0(z-z^3), \quad \rho_0 w_1 = \frac{\Ray}{24}\nabla\cdot\left(\frac{\rho_0}{\mu_0}\nabla\rho_0\right)(z^2-1)^2 \label{OBHR}
\end{equation}
The net discharge of this flow is zero, i.e., $\langle \vect v_0\rangle =0$. This would be non-zero in the presence of external factors (which are not considered in this paper) such as imposed pressure gradients such as in a Poiseuille flow or direct gravitational pressure gradients in inclined or vertical configurations. Equations~\eqref{OBHR} describe the leading-order solution for the flow field, provided $\rho_0$ and $\mu_0$ is known; the remaining analysis in this section will be devoted to deriving the governing equations for these quantities.

The flow described by the solution~\eqref{OBHR} is referred to as the Ostroumov--Birikh flow, primarily in Russian literature. However, it may also be referred to as Ostroumov--Birikh--Hansen--Rattray flows or simply Ostroumov flows\footnote{Ostroumov\cite{ostroumov1952svobodnaya} himself credits the solution to G. N. Guk in his textbook.}. The Ostroumov-type flows are distinct from the Poiseuille-type flows. In Poiseuille-type flows, $\vect v_0$ is a quadratic function of $z$ due to a constant pressure gradient at least on the local scale (i.e., $x,y\sim\ep$ and $z\sim 1$), although it may vary on the large scale (i.e., $x,y\sim 1$) in variable-density flows~\cite{rajamanickam2022effects,rajamanickam2023thick}. Moreover, we can deduce from~\eqref{OBHR} that since $\nabla\rho_0$ is constant, in the first approximation, on the local scale, $w_1$ is zero on the local scale, although it can vary on the large scale~\cite{chatwin1976some}. In contrast, in Poiseuille-type flows, $w_1$ is proportional to the first-derivative of $\rho_0$, leading to non-zero values on the local scale; see, for example,~\cite{pearce2014taylor,rajamanickam2022effects,rajamanickam2023thick}.

\subsection{First-order problem}
\label{sec:firstorder}
At the next order, we obtain
\begin{align}
\pfr{\rho_0}{t}+\nabla\cdot(\rho_1 \vect v_0 + \rho_0 \vect v_1) &= -\pfr{}{z}(\rho_1w_1+\rho_0 w_2), \label{cont1}\\
\frac{\rho_0  }{\Pra}\left(\vect v_0\cdot \nabla\vect v_0+w_1\pfr{\vect v_0}{z}\right) +\nabla p_1 &= \pfr{}{z}\left(\mu_0\pfr{\vect v_1}{z}+\mu_1\pfr{\vect v_0}{z}\right), \label{mom1}\\
\pfr{p_1}{z}&=-\Ray \rho_1, \label{zmom1}\\
   \rho_0 \vect v_0\cdot \nabla \theta_0 &= \pfr{}{z}\left(\lambda_0\pfr{\theta_1}{z}\right). \label{theta1}
\end{align}
The last three equations can be integrated using the boundary conditions~\eqref{Abc}-\eqref{flux} to determine $\vect v_1$, $p_1$ and $\theta_1$, which are presented in Appendix A. These solutions,  introduce three depth-averaged quantities, namely
\begin{equation}
    \vect V(\vect x,t) \equiv \langle \vect v_1 \rangle, \quad 3P(\vect x,t) \equiv \langle p_1\rangle, \quad \Theta(\vect x,t) \equiv \langle \theta_1\rangle.
\end{equation}
where $\vect V$ is given in~\eqref{meanV} and $\Theta$ is not needed in further analysis.
By an inspection of the continuity equation~\eqref{cont1}, we can also define an effective depth-averaged mass flux $\rho_0\vect u$ as
\begin{align}
      \rho_0\vect u = \langle\rho_0\vect v_1 + \rho_1\vect v_0\rangle = \rho_0 \vect V + \pfr{\rho_0}{\theta_0}\langle\vect v_0\theta_1\rangle  \quad \text{so that} \quad \pfr{\rho_0}{t}+\nabla\cdot(\rho_0\vect u)=0\label{effu}
\end{align}
ensuring $w_2(\vect x,\pm 1,t)=0$. The expression for $\vect u$ can be calculated using the solutions obtained so far and is presented below in~\eqref{DarcyA}.

\subsection{Second-order problem}
For the second-order problem, it is  sufficient to consider only the scalar equation~\eqref{thetadim} which simplifies to
\begin{align}
    \rho_0\pfr{\theta_0}{t} + (\rho_0 \vect v_1+\rho_1\vect v_0)\cdot\nabla\theta_0 + \nabla\cdot(\rho_0\vect v_0\theta_1) + \pfr{}{z}(\rho_0 w_1 \theta_1) \nonumber \\ 
    = \nabla\cdot(\lambda_0\nabla \theta_0)  + \pfr{}{z}\left(\lambda_0\pfr{\theta_2}{z}+\lambda_1\pfr{\theta_1}{z}\right) + Q_0(\vect x,t,\theta_0).  \label{T2}
\end{align}
The solvability condition for this equation, subject to the boundary condition $-\lambda_0\pfi{\theta_2}{z}=  \pm S_0^\pm(\vect x,t,\theta_0)$ at $z=\pm 1$, determines the  governing equation for $\theta_0(\vect x,t)$. Depth-averaging the above equation, the term $(\rho_0 \vect v_1+\rho_1\vect v_0)\cdot\nabla\theta_0$ on the left side becomes $\rho \vect u \cdot \nabla \theta_0$ (see~\eqref{effu}), whereas the term $\nabla\cdot(\rho_0 \vect v_0\theta_1)$ manifests as a buoyancy-induced diffusion term. Specifically, we find
 \begin{equation}
     \langle\vect v_0\theta_1\rangle =  - \gamma \Ray^2\frac{\rho_0\nabla\rho_0}{\mu_0^2\lambda_0} \nabla\rho_0\cdot\nabla\theta_0, \quad \text{where} \quad \gamma = \frac{2}{2835}.
 \end{equation}
 The $z$-derivative term on the left side of~\eqref{T2} vanishes upon averaging whereas the $z$-derivative terms on the right side incorporate contributions from the surface fluxes. All other terms in the equation remain unaffected by the averaging operation since they are independent of $z$. The resultant equation for $\theta_0(\vect x, t)$ thus obtained is presented in~\eqref{theta}.

\section{Two-dimensional depth-averaged governing equations}

Collecting \eqref{effu} and the depth-averaged version of~\eqref{T2}, we obtain the required two-dimensional depth-averaged governing equations. In writing down these equations, we shall omit the subscript $``0"$ for clarity. The governing equations then read
\begin{align}
\pfr{\rho}{t} + \nabla \cdot (\rho\vect u) &= 0, \label{cont}\\
-\mu \vect u = \nabla P &+ \mathbfcal F_\eff, \label{DarcyA} \\
       \rho\pfr{\theta}{t} + \rho \vect u\cdot \nabla \theta  &= \nabla\cdot(\rho \mathbfcal D_\eff \cdot\nabla\theta) +Q-S_\mathrm{avg}, \label{theta}\\  
      \rho =\rho(\theta), \quad \mu &= \mu(\theta) \quad \lambda=\lambda(\theta), \label{eqn}
\end{align}
in which $-S_\mathrm{avg} = -(S^++S^-)/2$ quantifies the average flux of scalar leaking through the walls and $\mathbfcal D_\eff$ is the effective diffusion matrix, which is defined by
\begin{equation}
   \rho\mathbfcal D_\eff =\lambda\left(\mathbf I +  \frac{\gamma \Ray^2\rho^2}{\mu^2\lambda^2}\nabla\rho\otimes\nabla\rho\right). \label{effD}
\end{equation}
The first part of the matrix represents normal molecular diffusion, whereas the second part represents buoyancy-induced diffusion. The matrix $\mathbfcal D_\eff$ is evidently symmetric and also positive definite as can be demonstrated; the term $\nabla \rho\otimes \nabla \rho$ is also known as the structure tensor or second moment matrix. The effective buoyancy force $\mathbfcal F_\eff$ that appears in Darcy's equation~\eqref{DarcyA} is given by
\begin{align}
  \frac{\mathbfcal F_\eff}{\gamma \Ray^2} =-\rho\nabla\left[\frac{(\nabla\rho)^2}{\mu\lambda}\right] -  \frac{3\rho\nabla\rho}{2\mu^2\lambda} \nabla\rho\cdot\nabla\mu + \frac{\rho(\nabla\rho\cdot\nabla)}{\mu\Pra}\left(\frac{\nabla\rho}{\mu}\right)    + \frac{3\nabla\rho}{2\mu\Pra}\nabla\cdot\left(\frac{\rho}{\mu}\nabla\rho\right). \label{effF}
\end{align}
It is crucial to note that the effective buoyancy force depends not only  on $\rho$, but also on $\mu$ and $\lambda$, although the later dependences vanish, as expected, in constant-density flows ($\nabla\rho=0$). 

Equations~\eqref{cont}-\eqref{effF} solve the unknown variables $\vect u$, $P$, $\theta$, $\rho$, $\mu$ and $\lambda$, subject to suitable initial condition and boundary conditions in the $xy$-plane. It is worth noting that the Rayleigh number appears in the equations only as $\Ray^2$ and as such the sign of the gravity vector, illustrated in Fig.~\ref{fig:sch}, is irrelevant in this leading-order problem. Of course, the leading-order flow field~\eqref{OBHR}, which is determined once~\eqref{cont}-\eqref{effF} is solved, does depend on the sign of the gravity vector.

As can be seen from~\eqref{effD}-\eqref{effF}, when $\Ray \sim \ep$, i.e., when the buoyancy-induced motion $U_b$ is comparable to the mixing speed $U_m$ as per the discussion made in section~\ref{sec:formulation}, the two effective quantities become $\rho\mathbfcal D_\eff=\lambda\vect I + O(\ep^2)$ and $\mathbfcal F_\eff = O(\ep^2)$. This suggests that both the shear-induced dispersion and buoyancy force disappear~\eqref{cont}-\eqref{effF}, recovering the classical equations including normal Darcy's law. 

The case in which density may be regarded as a variable physical quantity, but not the transport coefficients is of some interest. Then $\mu=\lambda=1$ and the effective diffusion matrix and buoyancy force simplify to 
\begin{align}
   \rho\mathbfcal D_\eff = \vect I + \gamma\Ray^2\rho^2\nabla\rho\otimes\nabla\rho, \quad \frac{\mathbfcal F_\eff}{\gamma \Ray^2} =\left(\frac{1}{2\Pra}-1\right)\rho\nabla(\nabla\rho)^2   + \frac{3\nabla\rho}{4\Pra}\nabla^2\rho^2. 
\end{align}

\subsection{Mutli-scalar system} \label{sec:multi}
Suppose that there are $N$ additional scalar fields $\varphi_i$, having a dimensional diffusion coefficient $D_i^*$, besides the scalar $\theta \equiv \varphi_0$.\footnote[3]{The subscripts $i$, including $0$, labelling different scalar fields should not confused with the subscripts used in the asymptotic analysis.} The density and the transport coefficients may, in general, depend on all $N+1$ variables, i.e.,
\begin{equation}
    \rho=\rho(\varphi_i), \quad \mu=\mu(\varphi_i), \quad \lambda=\lambda(\varphi_i), \qquad  i=0,1,2,\dots N
\end{equation}
and so are the volumetric source functions $Q_i= Q_i(\vect x,t,\varphi_i)$, and the average surface-flux functions $\tilde  S_{\mathrm{avg},i} =  S_{\mathrm{avg},i}(\vect x,t,\varphi_i)$. Defining $\Lew_i=D_i^*/D_0^*$ and assuming it to be constant, the equation for $\varphi_i$, in the absence of cross-diffusion between different scalar fields, can be written as
\begin{align}
    \rho\pfr{\varphi_i}{t} + \rho \vect u\cdot \nabla \varphi_i  &= \nabla\cdot(\rho\mathbfcal D_{\eff,i}\cdot\nabla\varphi_i) + {Q}_i- S_{\mathrm{avg},i}  \label{phi}
\end{align}
where 
\begin{equation}
  \rho \mathbfcal D_{\eff,i} = \frac{\lambda}{\Lew_i}\left(\mathbf I +  \frac{\gamma \Ray^2\Lew_i^2\rho^2}{\mu^2\lambda^2}\nabla\rho\otimes\nabla\rho\right). \label{effK}
\end{equation}
The formula for the effective buoyancy force~\eqref{effF} also needs modification with regards to those terms containing $\lambda$; specifically, we find
\begin{align}
    \frac{\mathbfcal F_\eff}{\gamma \Ray^2} & = - \rho\nabla\left[\frac{\nabla\rho}{\mu\lambda}\cdot\left(\sum_{i=0}^N\Lew_i \pfr{\rho}{\varphi_i}\nabla\varphi_i\right)\right]  -  \frac{3\rho\nabla\rho}{2\mu^2\lambda} \nabla\rho\cdot\left(\sum_{i=0}^N\Lew_i \pfr{\mu}{\varphi_i}\nabla\varphi_i\right)  \nonumber \\
    &\quad + \frac{\rho(\nabla\rho\cdot\nabla)}{\mu\Pra}\left(\frac{\nabla\rho}{\mu}\right)   + \frac{3\nabla\rho}{2\mu\Pra}\nabla\cdot\left(\frac{\rho}{\mu}\nabla\rho\right),  \label{MDarcy}
\end{align}
which reduces to~\eqref{effF} only for equi-diffusional system, i.e., for $\Lew_1=\Lew_2 = \dots=\Lew_N=1$; obviously, $\Lew_0=1$ always.

\section{Concluding remarks}

The development of an effective two-dimensional description, presented herein, is expected to pave the way for exploring a range of intriguing problems involving scalar transport in a Hele-Shaw cell. Future research can delve into canonical problems such as the evolution of non-axisymmetric hot spots and stability of horizontal convection. 

While the concept of shear-induced dispersion and its implications are well-established, the notion of a shear-induced force in Darcy's law is relatively novel, having been briefly mentioned in the previous work~\cite{rajamanickam2024effect}. The implications of this force are not immediately evident, as they depend on many specific physical quantities and may vary across different problems. 

Appendix B provides a useful exploration of the role of strong wall leakages in a specific case study. The challenge of investigating the impact of strong leakages under general conditions, as briefly mentioned in the footnote in Appendix B, warrants further investigation in future work. Extension of the current study to thin liquid layers, which are bounded above by a free surface, would also merit consideration in the future.

\section*{Acknowledgements}

The author is thankful to Joel Daou for some fruitful discussion regarding the problem addressed here. 

\section*{Appendix A: Solutions for the first-order problem}\label{appA}

Solutions to equations~\eqref{mom1}-\eqref{theta1} subject to~\eqref{Abc}-\eqref{flux} that arose in the first-order problem, are presented here. These are
\begin{align}
      \theta_1 &=  \Theta -  \frac{\Ray}{360}\frac{\rho_0}{\mu_0\lambda_0} \nabla\theta_0\cdot\nabla\rho_0 (3z^5-10z^3+15z) , \label{T1}\\  
      p_1 &=  3P - \Ray \pfr{\rho_0}{\theta_0}\left[\Theta z - \frac{\Ray}{5040} \frac{\rho_0}{\mu_0\lambda_0} \nabla\theta_0\cdot\nabla\rho_0(7z^6-35z^4+105z^2-29)\right], \label{P1}\\
      \mu_0\vect v_1 &=  \frac{3}{2}\nabla P(z^2-1) + \frac{\Ray}{6}\left[\nabla\left(\Theta\pfr{\rho_0}{\theta_0}\right)-\frac{\Theta\nabla\rho_0}{\mu_0}\pfr{\mu_0}{\theta_0}\right](z-z^3)  \nonumber \\
      &+\frac{\Ray^2}{120960}\nabla\left[\frac{\rho_0}{\mu_0\lambda_0}\pfr{\rho_0}{\theta_0}\nabla\theta_0\cdot\nabla\rho_0\right](z^2-1)^2(3z^4-22z^2+163) \nonumber  \\
      &-\frac{\Ray^2}{17280} \frac{\rho_0}{\mu_0^2\lambda_0}\nabla\rho_0 \pfr{\mu_0}{\theta_0}(\nabla\theta_0\cdot\nabla\rho_0)(z^2-1)(9z^6-35z^4+75z^2+15) \nonumber \\ &+\frac{\Ray^2}{30240\Pra} \frac{\rho_0}{\mu_0}\nabla\rho_0\cdot\nabla\left(\frac{\nabla\rho_0}{\mu_0}\right)(z^2-1)(15z^6-41z^4+29z^2+29)\nonumber\\
      & - \frac{\Ray^2}{120960\Pra} \frac{\nabla\rho_0}{\mu_0}\nabla\cdot\left(\frac{\rho_0}{\mu_0}\nabla\rho_0\right)(z^2-1)(45z^6-151z^4+199z^2-221), \label{v1}
\end{align}
in which $\Theta(\vect x,t)=\langle\theta_1\rangle$ and $3P(\vect x,t)=\langle p_1\rangle$ are integration constants, whereas $\vect V(\vect x,t)=\langle \vect v_1 \rangle$ is given by
\begin{align}
    \mu_0\vect V &=  -\nabla P + \gamma\Ray^2\nabla\left[\frac{\rho_0}{\mu_0\lambda_0}\pfr{\rho_0}{\theta_0}\nabla\theta_0\cdot\nabla\rho_0\right] + \frac{3}{2}\gamma\Ray^2\frac{\rho_0}{\mu_0^2\lambda_0}\nabla\rho_0 \pfr{\mu_0}{\theta_0}(\nabla\theta_0\cdot\nabla\rho_0) \nonumber \\
    & - \frac{1}{\Pra} \gamma\Ray^2\frac{\rho_0}{\mu_0}\nabla\rho_0\cdot\nabla\left(\frac{\nabla\rho_0}{\mu_0}\right) - \frac{3}{2\Pra}\gamma\Ray^2 \frac{\nabla\rho_0}{\mu_0}\nabla\cdot\left(\frac{\rho_0}{\mu_0}\nabla\rho_0\right) \label{meanV}
\end{align}
where $\gamma=2/2835$.

\section*{Appendix B: Role of strong leakages at the walls in a special scenario} \label{sec:strong}

The analysis carried out in the main text assumes that the flux leakage through channel walls at $z=\pm 1$ is small, as indicated in~\eqref{flux}. Let us explore how the results might change if the flux leakage is instead described by
\begin{equation}
    -\frac{\lambda}{\ep}\pfr{\theta}{z} =    R(\vect x,t,\theta) \quad \text{at} \quad z=\pm 1 \label{flux1}
\end{equation}
where $R$ is an order-unity function. This boundary condition pertains to a specific scenario where the average flux across the channel is zero. In other words, the flux entering one side of the channel is precisely balanced by the flux exiting the other side.\footnote[4]{A more general condition, $-(\lambda/\ep)\pfi{\theta}{z}= \pm R^\pm(\vect x,t,\theta)$ at $z=\pm 1$, is not considered here since it would generate a non-zero leading-order discharge, $\langle \vect v_0 \rangle \neq 0$, as implied by the solvability condition, $\rho_0 \langle \vect v_0 \rangle \cdot \nabla \theta_0 = -R_{0,\mathrm{avg}}$, of equation~\eqref{theta1}. This introduces a Poiseuille-type contribution to the leading-order Ostroumov flow~\eqref{OBHR}, indicating that the flow is not only driven by buoyancy forces, but also due to strong flux leakages. The primary challenge in incorporating non-zero values of $\langle \vect v_0 \rangle$ lies in the fact that the unknown function $\Theta=\langle \theta_1 \rangle$ would appear in the leading-order depth-averaged equations. This would create a closure problem, as we would need additional information to determine $\Theta$. Addressing this problem is beyond the scope of this work.}

With the revised condition for $\theta$, only the specific steps in the analysis from $\S$\ref{sec:asym} that require modifications will be highlighted. The first changes occur in the first-order problem discussed in \$\ref{sec:firstorder}. Equation~\eqref{theta1} for $\theta_1$ must now be supplemented with the boundary conditions $-\lambda_0\pfi{\theta_1}{z}= R_0$ at $z=\pm 1$. The resulting solution for $\theta_1$, $p_1$ and $\vect v_1$ is given by~\eqref{T1}-\eqref{v1} with the addition of the following three terms, respectively, to their right-hand side:
\begin{equation}
    - \frac{R_0}{\lambda_0} z, \qquad - \Ray \pfr{\rho_0}{\theta_0} \frac{R_0}{6\lambda_0}(1-3z^2), \qquad \frac{\Ray}{24}\nabla\left(\frac{R_0}{\lambda_0}\pfr{\rho_0}{\theta_0}\right)(z^2-1)^2.
\end{equation}
Turning to the second-order problem, we note that equation~\eqref{T2} is now subject to the boundary condition $-(\lambda_0\pfi{\theta_2}{z}+\lambda_1 \pfi{\theta_1}{z})= R_1$. When enforcing the solvability condition for equation~\eqref{T2}, we obtain a new contribution to the term $\langle\vect v_0\theta_1\rangle$, given by
 \begin{equation}
     \langle\vect v_0\theta_1\rangle =  - \gamma \Ray^2\frac{\rho_0\nabla\rho_0}{\mu_0^2\lambda_0} \nabla\rho_0\cdot\nabla\theta_0 - \gamma_s \Ray  \frac{R_0\nabla\rho_0}{\mu_0\lambda_0}, \quad \text{where} \quad \gamma = \frac{2}{2835}, \quad \gamma_s = \frac{1}{45}. 
 \end{equation}
The new constant $\gamma_s$ pertains to the contribution from the surface fluxes.

\textbf{Depth-averaged equations:} We can now write down the two-dimensional governing equations. As before, omitting the subscript $``0"$ for clarity, we find
\begin{align}
\pfr{\rho}{t} + \nabla \cdot (\rho\vect u) &= 0, \label{cont11}\\
-\mu \vect u = \nabla P &+ \mathbfcal F_\eff, \label{DarcyA11} \\
       \rho\pfr{\theta}{t} + \rho \vect u\cdot \nabla \theta   &= \nabla\cdot(\rho \mathbfcal D_\eff \cdot\nabla\theta) +Q, \label{theta11}\\  
      \rho =\rho(\theta), \quad \mu &= \mu(\theta) \quad \lambda=\lambda(\theta), \label{eqn11}
\end{align}
where
\begin{equation}
    \rho\mathbfcal D_\eff =\lambda\left(\mathbf I +  \frac{\gamma \Ray^2\rho^2}{\mu^2\lambda^2}\nabla\rho\otimes\nabla\rho\right) +  \gamma_s \Ray \frac{\rho R}{\mu\lambda} \pfr{\rho}{\theta} \mathbf I \label{Deff11}
\end{equation}
and
\begin{align}
 \mathbfcal F_\eff &=\gamma \Ray^2\left\{-\rho\nabla\left[\frac{(\nabla\rho)^2}{\mu\lambda}\right] -  \frac{3\rho\nabla\rho}{2\mu^2\lambda} \nabla\rho\cdot\nabla\mu + \frac{\rho(\nabla\rho\cdot\nabla)}{\mu\Pra}\left(\frac{\nabla\rho}{\mu}\right)    + \frac{3\nabla\rho}{2\mu\Pra}\nabla\cdot\left(\frac{\rho}{\mu}\nabla\rho\right)\right\},\nonumber \\
 & + \gamma_s\Ray\left[ \frac{R\nabla\rho}{\rho \lambda}\pfr{\rho}{\theta}- \nabla \left(\frac{R}{\lambda}\pfr{\rho}{\theta}\right)\right]. \label{effF11}
\end{align}
Thus, the primary modifications to the governing equations due to strong flux leakage through walls are the additional terms involving the function $R$ in the definitions of $\mathbfcal D_\eff$ and $\mathbfcal F_\eff$. These terms, proportional to $\Ray$, depend on the direction of gravity vector illustrated in Fig.~\ref{fig:sch}.

It is worth noting that although $\rho \mathbfcal D_\eff$ remains symmetric, it need not be positive definite. The potential for negative contribution arises from the product $\Ray\, R \pfi{\rho}{\theta}$. For most fluids, the expansion coefficient $-\rho^{-1}\pfi{\rho}{\theta}>0$ and therefore the aforementioned product becomes negative when $\Ray\, R>0$. As to when this condition is satisfied can be inferred from the following: (1) $\Ray>0$ when gravity points downwards as shown in Fig.~\ref{fig:sch} and $\Ray<0$ otherwise, and (2) $R>0$ when the net flux is downwards and $R<0$ when it is upwards.

\textbf{Multi-scalar system:} For the multi-scalar system discussed in $\S$\ref{sec:multi}, we now include  the boundary conditions $(-\lambda/\ep\Lew_i)\pfi{\varphi_i}{z}= R_i(\vect x,t,\varphi_i)$ at $z=\pm 1$. The formula for the effective buoyancy force $\mathbfcal F_\eff$ is then obtained by adding, to the right-hand side of~\eqref{MDarcy}, the term,
\begin{equation}
    \gamma_s \Ray \left[\frac{\nabla \rho}{\rho\lambda}\sum_{i=0}^N \Lew_i  R_i \pfr{\rho}{\varphi_i}-\nabla \left(\frac{1}{\lambda}\sum_{i=0}^N \Lew_i  R_i \pfr{\rho}{\varphi_i}\right)\right].
\end{equation}
The governing equation~\eqref{phi} for $\varphi_i$ must now be modified due to the emergence of cross-diffusion terms. The modified equations are given by
\begin{align}
    \rho\pfr{\varphi_i}{t} + \rho \vect u\cdot \nabla \varphi_i  &= \nabla\cdot(\rho \mathbfcal D_\eff \cdot\nabla\varphi_i) + \sum_{j=0,j\neq i}^N\nabla\cdot (\rho \mathbfcal D_{c,\eff,j}\cdot \nabla\varphi_j) \,+Q_i 
\end{align}
where
\begin{align}
 \rho\mathbfcal D_{\eff,i} &= \frac{\lambda}{\Lew_i}\left(\mathbf I +  \frac{\gamma \Ray^2\Lew_i^2\rho^2}{\mu^2\lambda^2}\nabla\rho\otimes\nabla\rho\right) +  \gamma_s \Ray \Lew_i\frac{\rho R_i}{\mu\lambda} \pfr{\rho}{\varphi_i} \mathbf I, \nonumber \\
    \rho \mathbfcal D_{c,\eff,j} &= \gamma_s \Ray \Lew_i\frac{\rho R_i}{\mu\lambda}\pfr{\rho}{\varphi_j}\vect I.
\end{align}

\textbf{Remarks for extremely strong leakages:} We define the extremely strong  leakage as the condition where $-\lambda \pfi{\theta}{z}= T^{\pm}(\vect x, t,\theta)\sim O(1)$ at $z=\pm 1$ holds. This implies that $\theta$ varies significantly on the scale $z\sim 1$  (or $z^*\sim h$ in dimensional units), rendering the notion of a thick mixing region less relevant. In such cases, the problem becomes fundamentally three-dimensional, and deviations from the depth-averaged value are expected to be of order one. Moreover, the flow induced by these extremely strong leakages will also be exceptionally strong, i.e., $\vect v \sim O(1/\ep^2)$ and $p\sim O(1/\ep^2)$. This means that the leading-order flow is no longer driven by buoyancy forces but rather by density variations, such as thermal expansion, imposed by the intense leakages.

\bibliographystyle{unsrt}
\bibliography{references}

\end{document}